# A Machine Learning Framework for Extending Wave Height Time Series Using Historical Wind Records

**Hazem U. Abdelhady, Cary D. Troy***

Purdue University, Lyles School of Civil Engineering, 550 Stadium Mall Drive, West Lafayette, IN, 47907-2051, USA

\* Corresponding author: troy@purdue.edu

## Abstract

This study presents a novel machine learning-based (ML) framework that utilizes the ConvLSTM-1D model to hindcast or forecast wave heights at coastal locations using a nonuniform array of wind observations. This approach was applied to Lake Michigan to perform a 70-year ice-free hindcast of waves near Chicago, IL (USA). The Wave Information System model (WIS) served as the training, validation, and testing dataset for the ML model. Ensemble learning-optimized ML models forced by different numbers of observation stations were tested, showing that a single wind station alone as an input feature produced a reasonably accurate wave height model. However, the wave height model accuracy increased as more wind input data was included from around the lake, largely plateauing beyond the inclusion of four stations that spanned Lake Michigan's southern basin. The optimized model lookback period was found to be 10 hours for all models, suggestive of a fetch-limited temporal coupling between the wind observations and nearshore waves. The ML framework offers a promising avenue for utilizing historical wind records worldwide to extend wave height time series for nearshore locations, particularly in enclosed and semi-enclosed basins where waves are strongly linked to local winds.

## Introduction

Accurate measurement and simulation of nearshore wave characteristics play a crucial role in various fields, particularly coastal engineering, marine operations, and environmental research. Wave height, in particular, is the key characteristic used in the design and assessment of coastal structures such as breakwaters, seawalls, and harbors (Ciria et al., 2006). Furthermore, wave height time series serve as a valuable input for environmental research and monitoring programs, aiding in the calculating of different coastal processes (Dean & Dalrymple, 2001), predicting shoreline changes (Abdelhady & Troy, 2023), assessing climate change impacts (Semedo et al., 2012), and developing sustainable management strategies for coastal ecosystems (Gracia et al., 2018).

Long-term wave height records and hindcasts capture the variability, trends, and extremes in wave climate over extended periods, offering invaluable insights into the behavior of nearshore coastal areas (X. L. Wang & Swail, 2006). By analyzing long-time series of wave heights, researchers and engineers can identify and quantify long-term changes in wave patterns, including shifts in wave height distributions, wave frequency, and storm intensities (Guedes Soares & Scotto, 2004; X. L. Wang & Swail, 2006). This information is crucial for understanding the impacts of climate change on coastal regions and evaluating the vulnerability of coastal infrastructure (Guedes Soares & Scotto, 2004). Furthermore, long-time series of wave heights provide a basis for designing and optimizing coastal structures, as they enable engineers to more accurately estimate the statistical properties of waves needed for design, such as extreme wave heights corresponding to particular return periods (Guedes Soares & Scotto, 2001; Reeve et al., 2016).

Despite the importance of having long duration wave records for assessing coastal wave climates, most coastal areas have wave information spanning several decades at most. As an example, the U.S. Army Corps of Engineers Wave Information System (WIS) provides the longest wave records available for Lake Michigan (USA), spanning a period from 1979 to 2021, amounting to only 42 years of wave information (U.S. Army Corps of Engineers, n.d.). This limited temporal coverage hampers the ability to fully understand and predict long-term wave behavior in these areas and leads to substantial uncertainty in estimated extreme wave heights used for design, which often have return periods exceeding the record duration.

Physics-based numerical models have long been the primary approach for hindcasting and forecasting wave heights in coastal areas. These models employ physics-based equations that predict wave characteristics from various factors including the spatial-temporal wind distribution, water depths, bottom composition, and currents (also simulated). Among the most used physics-based models today are WaveWatch III (Tolman et al., 2002) and SWAN (Booij et al., 1999), which have been applied extensively to provide valuable information for a wide range of locations and applications. However, because these models require well-resolved wind fields (in space and time), their application for historical periods becomes challenging for locations and/or periods where wind data is sparse in space and/or time. The requirement of spatially-distributed wind fields is an unfortunate limitation, because discrete wind observations near coastal sites are available from as early as the late 1800s, and can conceivably be leveraged for historical wave prediction. Physics-based wave models can also be computationally expensive.

Machine learning (ML) models offer one efficient possibility for using wind observations to extend wave simulation records farther into the past when wind observations are sparse. ML models have been used for predicting the wave height either at specific locations (Deo et al., 2001; Hu et al., 2021; Kumar et al., 2018) or for a particular region using multiple points or a grid-based approach (Feng et al., 2020; James et al., 2018; Song et al., 2022; Zilong et al., 2022). Most of these studies either focus on the use of ML as a computationally efficient tool for hindcasting wave heights (Feng et al., 2020; James et al., 2018; Zilong et al., 2022), or as an efficient tool to predict future wave heights from past conditions (Song et al., 2022; S. Zhou et al., 2021).

The ML algorithms used in these papers can be broadly classified into two main categories: (1) non-recurrent machine learning techniques (e.g., Multi-Layer Perceptron (MLP), Supported Vector Machine (SVM), random forest); and (2) recurrent neural network (RNN) models (e.g., Long Short-Term Memory (LSTM), Convolution LSTM (ConvLSTM)). When applied to wave height prediction, non-recurrent models utilize input data available at a specific time step to predict wave height at the same time step, while RNN models account for the temporal correlation by considering both current and previous time steps. A summary of these models, along with their respective advantages and disadvantages for wave prediction, is presented in Table 1. Because the wave state at any point in time results from the wind field experienced in the past, accounting for temporal correlations between waves and winds through the RNN model framework can improve the simulation accuracy (Song et al., 2022) as well as the model's ability to solve more complex systems (James et al., 2018).

*Table 1 Comparison between different machine learning models that have been used for wave prediction.*

| Method | Type | Description | Advantages | Disadvantages | Applications |
|---|---|---|---|---|---|
| Supported Vector Machines (SVM) | Non-recurrent | Curve-fitting technique | • Less prone to overfitting<br>• Good with small training data | • Does not account for temporal correlations between wind and waves<br>• Less suitable for noisy data | (Berbić et al., 2017; Malekmohamadi et al., 2011) |
| Random Forest | Non-recurrent | Ensemble of decision trees | • Easy to implement<br>• Provides features importance ranking<br>• Less prone to overfitting | • Not good with complex nonlinear mappings<br>• Does not account for the temporal correlation | (Callens et al., 2020) |
| Multi-Layer Perceptron (MLP) | Non-recurrent | Feedforward neural network | • Can handle complex nonlinear mapping | • Does not account for the temporal correlations<br>• Not the best in capturing spatial correlations | (Deo et al., 2001; Feng et al., 2020; James et al., 2018) |
| LSTM | Recurrent models | A type of recurrent neural network (RNN) | • Account for the temporal correlation | • Not the best in capturing spatial correlations | (Fan et al., 2020; Hu et al., 2021; Zilong et al., 2022) |
| ConvLSTM | Recurrent models | Combines Convolution Neural Networks (CNNs) with LSTM | • Account for spatial correlation<br>• Account for the temporal correlation | • Harder to train than LSTM<br>• More prone to overfitting<br>• Requires 2D grided data | (Ouyang et al., 2023; Song et al., 2022; S. Zhou et al., 2021) |
| ConvLSTM 1D | Recurrent models | Combines Convolution Neural Networks (CNNs) with LSTM | • Account for spatial correlation<br>• Account for the temporal correlation | • Harder to train than LSTM<br>• More prone to overfitting<br>• Requires 1D grided data | This study |

One potential advantage of machine learning models for the prediction of locally-generated waves in coastal areas is that they can be driven exclusively by land-based wind observations if water observations are unavailable. However, care must be taken in selecting an appropriate ML model that can successfully leverage land observations. Deo et al. (2001) showed that utilizing a single onshore wind station to predict wave height at a buoy location significantly reduced the accuracy of wave height predictions compared to using an offshore wind station at the location of interest for the wave height. This discrepancy in accuracy can be attributed to the temporal mismatch between the occurrence of storms at the wind station and its impact at the wave station, which was not adequately accounted for in the Multi-Layer Perceptron (MLP) algorithms used in the study. These findings point to the importance of developing methodologies that effectively utilize multiple available onshore wind records to enhance the precision of wave height predictions in nearshore coastal areas, which is the primary focus of this study.

Various ML models are available to simulate waves from wind data (Table 1), and the selection of the ML model is an important consideration. For the present application, historical wind observation stations that are located on land at varying distances and directions from the wave simulation location are used. From a modeling perspective, each wind station has a unique relationship between the wind velocities and directions at the station and the waves at the simulation site; additionally, these relationships have a temporal component related to the time differences between winds and waves. The chosen ML algorithm for this study should ideally be capable of capturing and leveraging these multiple, complex spatial and temporal wind-wave relationships, in order to achieve accurate predictions for wave heights.

As mentioned earlier, temporal dependence between meteorological data and wave data can be effectively captured using Recurrent Neural Networks (RNNs). Among various RNN algorithms, Long-Short Term Memory (LSTM) is widely employed for time series analysis, typically to predict future time steps in the series from previous time steps. To use it for multiple outputs, (Sutskever et al., 2014) introduced sequence-to-sequence learning by stacking multiple LSTM together, which can be used to map the temporal correlation between multiple inputs and multiple outputs. Although this technique is good for dealing with multiple sequences, it is not ideal for capturing the spatial correlation between the sequences (Shi et al., 2015).

To address this limitation, (Shi et al., 2015) introduced the ConvLSTM algorithm in the context of precipitation forecasting, building upon the idea of stacked LSTM while incorporating convolution operators in the state-to-state and input-to-state transitions. Unlike the standard Hadamard product used in stacked LSTMs, ConvLSTM employs convolution operators to overcome the spatial correlation limitation. For a comprehensive understanding of the ConvLSTM's key equations, please refer to the original paper (Shi et al., 2015).

The ConvLSTM algorithm has demonstrated promising accuracy in predicting multiple geophysical factors as it can better capture the spatial and temporal correlation between the inputs and the outputs (Reichstein et al., 2019; Ren et al., 2018; Shi et al., 2015). Furthermore, it has been successfully implemented in several recent studies to forecast future wave heights from meteorological data (Ouyang et al., 2023; Song et al., 2022; S. Zhou et al., 2021).

The present work develops and tests a machine learning model framework for wave prediction that utilizes historical wind records from multiple land-based observation locations. The model is refined and applied to generate hindcasts of wave heights at a particular location in Lake Michigan, one of the Laurentian Great Lakes, near Chicago (USA). By employing the ConvLSTM-1D algorithm, the model establishes a mapping between land-based wind data and wave height data, attempting to capture both spatial and temporal correlations between observed wind records and the waves at the simulation location. As input data to drive the wave predictions, the model utilizes onshore Automatic Surface Observation Systems (ASOS) wind stations with historical wind records dating back to the 1940s, which is significantly earlier than most existing wave hindcasts. Multiple models were created and tested to determine the optimal number of stations for accurate wave height predictions and optimized model hyperparameters. The training, validation, and testing of the models were conducted by comparing the model prediction to the Wave Information System (WIS) wave heights. Ensemble learning was used to combine different models to obtain an ensemble model with better accuracy and generalization abilities than any individual model. Subsequently, the best-ensemble model was utilized to hindcast wave heights from 1949 to 1979, enabling the construction of a comprehensive wave height time series for Chicago spanning from 1949 to the present.

## Methods

### Study site

The nearshore site for the ML wave modeling undertaken in this study is located near Chicago, IL (USA) in Lake Michigan at 41.88°N, 87.56 °W (Figure 1). Lake Michigan is one of the five Laurentian Great Lakes, with more than 2,500 km of coastline, a maximum depth of 281 m, and an average depth of 85 m (NCEI, 1996). The lake is elongated in the north-south direction, with a length of approximately 500 km, and this large fetch allows strong autumn and winter storms to create spectral significant wave heights that can exceed 6 m, with corresponding dominant periods of 12-13 sec (e.g. https://www.ndbc.noaa.gov/station_page.php?station=45007 ). Long period swell (>15 s) is not present due to the enclosed geometry of the lake, and waves are generated entirely by local winds experienced on the lake. This feature differs from open ocean coasts for which local winds can be decoupled from the longer period waves in the spectrum. Lake Michigan does not experience astronomical tides of any dynamical consequence, but the lake water levels can fluctuate by up to 2 m on 5-10 year timescales due to variations in the hydrologic cycle, and these water level fluctuations drive large shoreline changes (Troy et al., 2021).

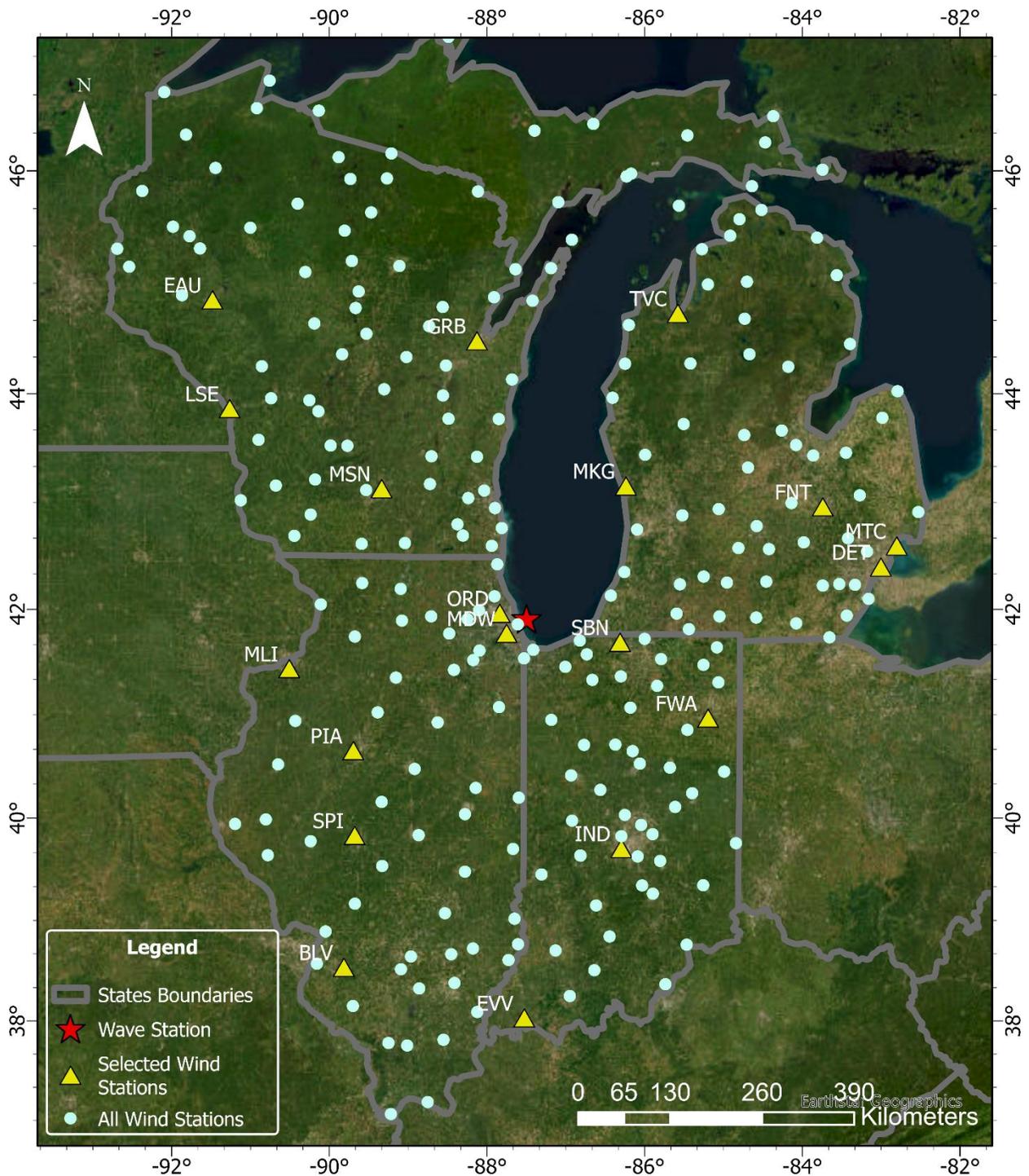

*Figure 1: Available ASOS wind observation stations around Lake Michigan, selected wind stations based on long-term data availability, and the simulated wave location near Chicago, IL.*

Input wind data

Land-based wind observations were used as the input features for the ML wave model developed in this study. Automatic Surface Observation Systems (ASOS) stations were used as the source of the wind data. The ASOS stations data was downloaded from

https://mesonet.agron.iastate.edu/request/download.phtml (accessed on 8/22/2022). Figure 1 shows the 256 stations found in the four states around Lake Michigan. However, many of these stations were not used in this study either because they are very recent stations, do not cover the period of interest from 1949 to 2020, or have extended, continuous data gaps that exceed one year. Out of the 256 stations, 19 stations matched these two criteria (Figure 1). From these data, wind velocities in the east and north directions were derived from the wind speed and direction to be used as the input features for the ML model.

While the ASOS data are quality-controlled, additional manual and automated quality control was performed on the wind speeds to remove clearly erroneous data points. Additionally, datasets containing sub hourly values were resampled to hourly time steps. Then, from all stations of hourly east and west wind speeds, simulation time steps were determined as times for which data from all stations (for a model run with a given collection of stations) were available. Any time steps with missing data points from the stations used in the training were subsequently excluded from the dataset. The number of gaps in the combined dataset increases with the number of stations used. The issue of input data gaps is discussed later in the paper.

Principal Component Analysis (PCA) was employed as an initial step in analyzing the wind data after preprocessing. PCA is a widely used statistical technique that enables dimensionality reduction in large datasets or datasets with numerous features (Witten et al., 2016). By applying the PCA, a smaller number of features, the principal components, can sometimes be used instead of the main features, which may streamline and improve the model performance and its generalization ability. It can also ascertain whether the chosen wind stations provided new and distinct information or if they merely duplicated existing information.

## Wave information

Wave data is available from buoys throughout the Great Lakes from the NOAA National Buoy Data Center (https://www.ndbc.noaa.gov/). However, due to winter ice cover, all Lake Michigan buoys are removed from the lake during the winter period, which leaves significant data gaps during the portion of the year when waves are generally largest. To overcome this data limitation, simulated wave heights from the United States Army Corps of Civil Engineers (USACE) Wave Information Studies (WIS) wave model was used to train and validate the ML wave model in this study. The wave variable simulated was the spectral significant wave height, Hm0. The WIS model provides long-term wave climatology for all of the United States coastal areas, including the Great Lakes and Lake Michigan, the focus of this study (U.S. Army Corps of Engineers, n.d.). This model has been refined and validated using available buoy observations for Lake Michigan (Jensen et al., 2012), and in the present modeling effort serves as a continuous data surrogate for ML model training. The WIS model output can be downloaded from https://wisportal.erdc.dren.mil/ (accessed on 3/23/2023). In this study, the simulated significant wave height from a single location, station ST94015 in Lake Michigan near Chicago was used, for the period 1979-2021 (Figure 1).

The WIS dataset was divided into three distinct datasets for training, validation, and testing (we hereafter refer to the wave heights simulated by WIS as "data", because it is used as such in the training and validation of the ML model developed herein). The training dataset spanned from 1989 to 2021 and served as the foundation for training the ConvLSTM models developed in this study. The validation dataset covered the period from 1979 to 1984, and was used in fine-tuning the model's hyperparameters to enhance the model's accuracy and generalization ability in addition to choosing the optimum number of stations and the best ensemble model as discussed later in the paper. Finally, the testing dataset extended from 1984 to 1989, was used as a final blind testing phase to evaluate the final model's performance and to provide an unbiased assessment of the selected model's ability to predict wave heights on unseen data.

Ice cover in Lake Michigan can significantly affect the waves during the winter in the Great Lakes and near Chicago. The effect of ice cover occurring over the wave fetch or at a particular location is to diminish wave heights. Following guidance from (Anderson et al., 2015; Feng et al., 2020; Tuomi et al., 2011), wave heights for days with local ice cover of more than 30% were set to zero for the final time series, with ice cover data obtained from the NOAA Great Lakes Ice Atlas for the period of 1973 to the present (https://www.glerl.noaa.gov/data/ice/). The effect of ice on the ML model is discussed further in the Discussion section of the paper.

## Machine learning model for wave simulation

Despite the ConvLSTM model's ability to capture spatial and temporal correlations, its usage for wave prediction thus far has involved only situations where input wind data is from gridded over-water locations in the immediate vicinity of the wave simulation site (e.g., (Feng et al., 2020; S. Zhou et al., 2021)) This study aims to predict wave heights at a coastal location without co-located wind measurements, from a randomly distributed network of land measurement stations.

To address these differences, the 1D version of the ConvLSTM model, ConvLSTM-1D, was selected for this study as it is easier to arrange the wind data into a 1D array than in a 2D array and cheaper computationally. Then, the output of the ConvLSTM-1D was flattened and passed to a dense layer with 1 output that represents the wave station (Figure 2). The ConvLSTM-1D in this study was implemented in the Python library, Keras (Chollet & others, 2015).

To apply the ConvLSTM-1D model, it was necessary to first construct a 1D grid comprising the ASOS wind stations, as these stations are not inherently gridded. Various permutations of the station's location within the grid are possible. In this study, all the stations were arranged and added to the 1D grid based on their proximity to the wave station being simulated. The closest station was positioned at the far left, while the farthest station was placed at the far right (refer to Figure 2). This sorting approach was chosen to facilitate adding or removing stations from the grid to be able to determine the optimum number of stations as discussed later. The effect of an alternative arrangement of the stations within the 1D grid and its impact on the model's performance is further explored and discussed in the later sections of this paper.

One of the key assumptions underlying LSTM-based models is the presence of a constant time step in the input data. While this assumption holds for most of the dataset used in this study, there are instances where wind data gaps exist and failure to account for these variable gaps can compromise the model's accuracy. These gaps may arise from missing data points at specific stations, data removal during the data quality check process, or ice conditions. To address this challenge, a time feature was introduced to the 1D grid, representing the time difference between consecutive time steps to give the model the ability to discern between regular and irregular time steps (Figure 2). For example, under normal circumstances with an hourly time step, this feature would have a value of 1 hour. However, in the presence of missing time steps, it would indicate the number of missing time steps.

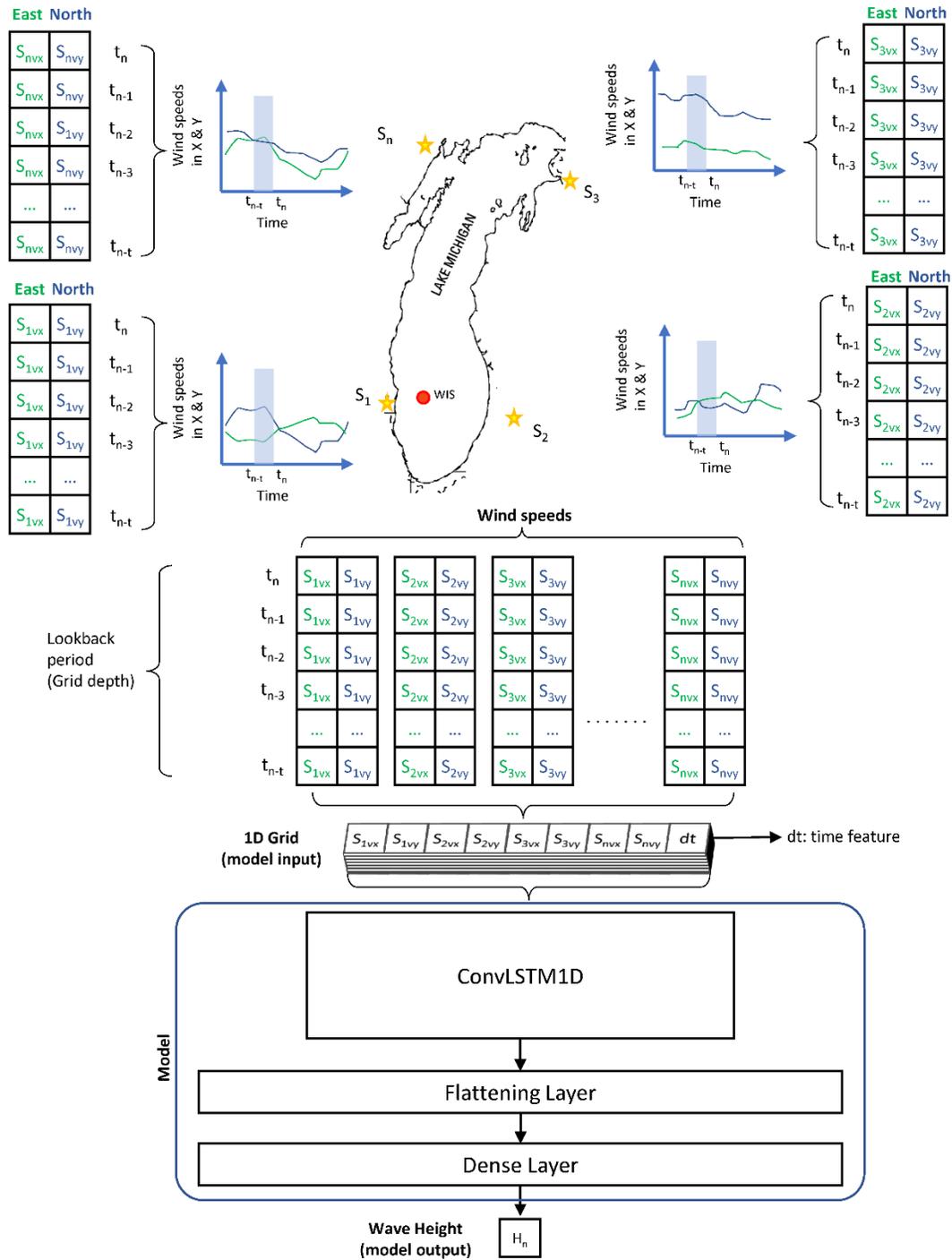

*Figure 2 Schematic showing the steps needed to predict the wave height from one 1D-ConvLSTM model based on the ASOS wind station input data.*

The ConvLSTM-1D contains several hyperparameters that were optimized to improve and refine model performance for this study. These parameters are briefly introduced here and optimized values are discussed later in the manuscript. One important hyperparameter that was optimized was the kernel size. The kernel size governs the size of the sliding window during convolution, impacting the scope of the spatial context considered (Goodfellow et al., 2016). In the context of wave prediction from discrete wind stations, selecting an appropriate kernel size is crucial for aggregating data from different wind stations effectively, capturing their spatial correlations. Balancing between local patterns and broader spatial trends, the kernel size significantly influences the model's ability to learn and predict wave heights accurately based on the provided wind station information.

Another main hyperparameter that was fine-tuned in the ConvLSTM-1D models was the number of filters. Filters are the learnable parameters applied in the ConvLSMT-1D model during convolution operations. The number of filters directly shapes model depth and output channels, impacting the model's complexity and learning capacity (Goodfellow et al., 2016). A large number of filters enhances the model's ability to capture complex spatial relationships from scattered wind station data but raises overfitting and training complexity risks. Conversely, fewer filters yield a simpler model, fitting limited data better, and reducing overfitting potential.

Another essential hyperparameter in LSTM-based models is the lookback period. The lookback period determines the number of preceding wind time steps utilized to predict the subsequent wave height which is important as waves at a given time are a function of the recent wind history, not just the most recent winds experienced. Increasing the lookback period enhances the model memory and increases the model complexity but results in a harder-to-train model and more prone to overfitting, while decreasing the lookback period results in a faster and easier-to-train model yet increases the potential loss of long-term trend (Goodfellow et al., 2016).

## Optimum model selection

In this study, numerous ConvLSTM-1D models were trained using various datasets, with each dataset having a different number of wind stations, in order to determine the optimal number of wind stations required to accurately predict wave heights at the target location near Chicago. The wave height models' weights were trained using backpropagation of the mean square error (MSE) loss with the Adam optimizer (Kingma & Ba, 2014). The models used data from a single wind station initially and progressively included additional stations up to a maximum of 19 stations (Figure 3). The selection of stations for each dataset was based on their proximity to the predicted wave height location (Table 2). Consequently, the first dataset incorporated the closest ASOS station (MDW), while subsequent datasets included the closest two ASOS stations, and so forth (see Table 2 for the ordered station list).

While the training, validation, and testing datasets for all models using different numbers of wind stations share the same starting and ending dates, it is important to note that the length of these datasets varies across different models. This discrepancy arises due to the inclusion of only those time steps that are available at all stations. Consequently, as the number of stations increases, there is a higher likelihood of encountering more missing time steps, leading to a reduction in the overall dataset size. For example, the length of the training dataset ranges between 332,983 and 227,584 time steps for the 1-station and 19-station models, respectively, for the same period between 1989 and 2021 (Table 2).

*Table 2 ASOS stations used to in the models*

| Station number | Station name | Location | Distance from wave station (km) | Number of observations available for combined model |
|---|---|---|---|---|
| 1 | MDW | Chicago, IL | 26 | 332,983 |
| 2 | ORD | Chicago, IL | 34 | 329,136 |
| 3 | SBN | South Bend, IN | 142 | 324,930 |
| 4 | MKG | Muskegon, MI | 243 | 313,461 |
| 5 | MSN | Dane County, WI | 274 | 304,545 |
| 6 | FWA | Fort Wayne, IN | 296 | 302,479 |
| 7 | PIA | Peoria County, IL | 299 | 297,187 |
| 8 | MLI | Rock Island County, IL | 334 | 293,502 |
| 9 | IND | Indianapolis, IN | 349 | 291,254 |
| 10 | SPI | Sangamon County, IL | 382 | 288,725 |
| 11 | GRB | Brown County, WI | 403 | 286,485 |
| 12 | FNT | Genesee County, MI | 456 | 283,379 |
| 13 | TVC | Traverse City, MI | 490 | 267,133 |
| 14 | LSE | La Crosse, WI | 511 | 260,304 |
| 15 | DET | Detroit, MI | 514 | 254,855 |
| 16 | MTC | Macomb County, MI | 541 | 242,839 |
| 17 | BLV | St. Clair County, IL | 548 | 233,338 |
| 18 | EVV | Evansville, IN | 558 | 230,319 |
| 19 | EAU | Chippewa County, WI | 631 | 227,584 |

As such, even though the inclusion of more wind stations increases the information available to train the models, this improvement is somewhat offset by the reduced data available to train the models with more stations. This characteristic renders models with more stations more complex, posing challenges in terms of training and increasing their susceptibility to overfitting. These factors necessitate careful consideration and the implementation of appropriate strategies to mitigate the complexities associated with models incorporating a greater number of stations like early stopping and dropout layers. Early stopping is a technique used to prevent overfitting by monitoring the model's performance on a validation dataset and stopping the training process when the performance starts to degrade (Yao et al., 2007). Dropout layers, on the other hand, are a regularization technique that randomly drops a fraction of the neurons during training, forcing the model to learn more robust and generalizable features (Hinton et al., 2012).

Since different models with different numbers of stations have different numbers of features and different complexity levels, the kernel size and the number of filters should change from one model to the other. Different combinations of kernel size, number of filters, and lookback periods were tested during the training period using the grid-search technique to determine their effect and to choose the best combination for each of the models.

For each of the 19 combined wind datasets modeled, 20 ConvLSTM-1D models were trained to ensure the consistence of the training process and to be used for ensemble learning (Figure 3). Ensemble learning is a powerful approach in machine learning that involves combining different models to create a final model that surpasses the accuracy of individual models (Zhou, 2009). In this study, the trained models for a certain dataset were ranked based on the mean square error of the simulated wave heights (MSE), and their predictions were algebraically summed to form the ensemble models. Consequently, a series of ensemble models were created, ranging from 2 models to 20 models, to achieve superior accuracy and predictive capability compared to any individual model used independently.

Consequently, the comparison between different ensemble models was done on the validation datasets. Ensemble model with the best MSE was selected and blindly tested on the testing dataset. The performance metrics, discussed later, were used to compare the change in the ensemble model's performance between the validation and the testing datasets (Figure 3). Agreement between the ensemble model performance on both datasets can ensure its reliable performance and generalization abilities and avoids overfitting the validation dataset.

Finally, the best ensemble model was selected to generate an ice-free time series of wave heights for the whole record of wind speeds. Then, the ice cover effect was implemented by excluding days with ice cover more than 30% as mentioned earlier to get the final time series (Figure 3).

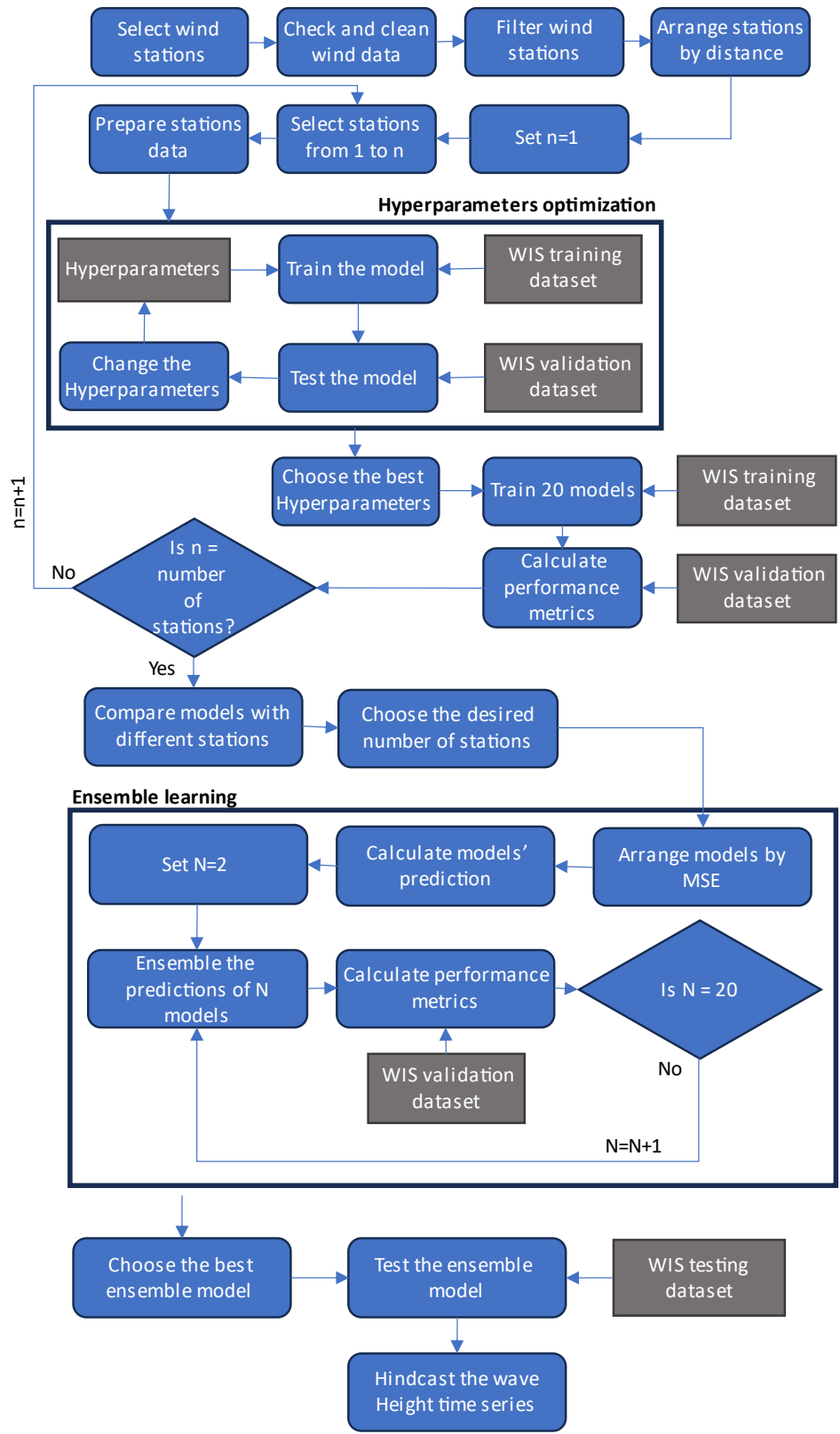

*Figure 3 Flow chart for the overall ML modeling framework, from wind station selection to hindcasting the wave height time series*

Performance metrics

Several metrics were utilized to assess model performance and facilitate comparisons and optimizations between different models. The mean square error (MSE) quantifies the average squared difference between predicted and actual (WIS-simulated) wave heights which was also used as the loss function for the model training. Bias measures the systematic deviation between predicted and actual values. The correlation coefficient evaluates the strength and direction of the linear relationship between predicted and actual values, while R2 (coefficient of determination) gauges the proportion of variance in the dependent variable explained by the independent variables. These analyses allowed for effective model comparisons and the selection of the most accurate and reliable models for predicting wave heights.

Standard definitions of MSE, bias, and R2 were used to quantify model performance. To calculate MSE, the squared differences between predicted and actual values were averaged. Bias was determined by taking the average difference between predicted and actual values. The correlation coefficient was calculated using statistical methods like Pearson's correlation coefficient, measuring the degree of linear relationship. R2 is calculated by subtracting the ratio of the sum of squared residuals from the total sum of squares from 1.

Results

Performance (validation) metrics for the optimized models, as a function of the number of wind stations used for each optimized model, are shown in Figures 4, 5, and 6. Based on the metrics used to assess the models, all models show very good agreement with the wave validation data, even the model that only utilizes data from a single wind station (Midway International Airport, MDW, the closest station to the simulation site). The MDW-only model has a mean squared error of 0.043 $m^2$, a bias of -0.022 m, and a correlation coefficient of 0.71. These values are comparable to the model performance of the WIS model itself, when compared to direct observations of wave heights in Lake Michigan (Jensen et al., 2012). This result suggests that if enough wind data is available from a single location near the wave simulation site, the ConvLSTM-1D model can be successfully trained to simulate wave heights. However, it should be noted that in case of one station, the ConvLSTM-1D model does not provide any advantage over the normal LSTM model.

As more wind stations are added to the model, the ML model performance improves, as would be expected. Figures 4, 5, and 6 show that there is substantial model improvement through the inclusion of the four nearest observation stations, after which the model does not improve appreciably with the addition of stations. The most accurate model (relative to the validation wave heights from WIS) is the model that utilizes the 13 closest observation stations to the Chicago simulation site. This "best" model has a MSE of 0.022 $m^2$, a negligible bias, and a correlation coefficient of 0.85. It should be noted, however, that the performance of the 13-station model is only marginally better than the 4-station model, which has MSE of 0.024 $m^2$, a bias of -0.004, and a correlation coefficient of 0.84. The 4-station model includes the wind stations of MDW, ORD, SBN, and MKG (Table 2).

The 4-station and 13-station models each possess distinct advantages. The 4-station model demonstrates slightly lower accuracy but requires less data and training time. Conversely, the 13-station model achieves slightly higher accuracy but requires more data and a longer training time. Moreover, models with more stations have more missing data for training as discussed earlier. Accordingly, the results of the ensemble learning stage is focused on these two models as well as the most basic 1-station model.

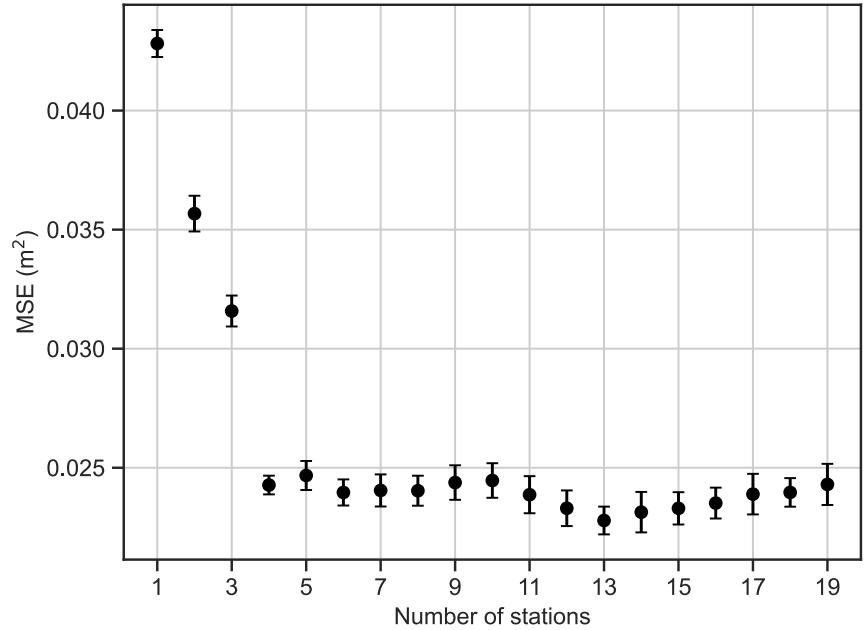

*Figure 4 MSE of all the optimized models using different number of wind stations. The error bars indicate +/- one standard deviation of the 20 models' MSE.*

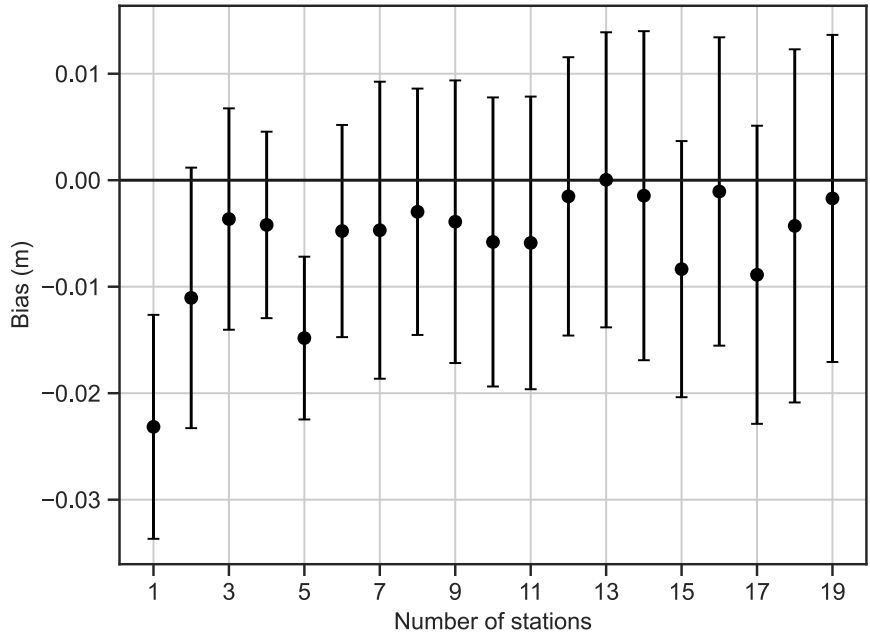

*Figure 5 The modeled wave height bias of all the models using different datasets. The error bars indicate +/- one standard deviation of the 20 models' bias.*

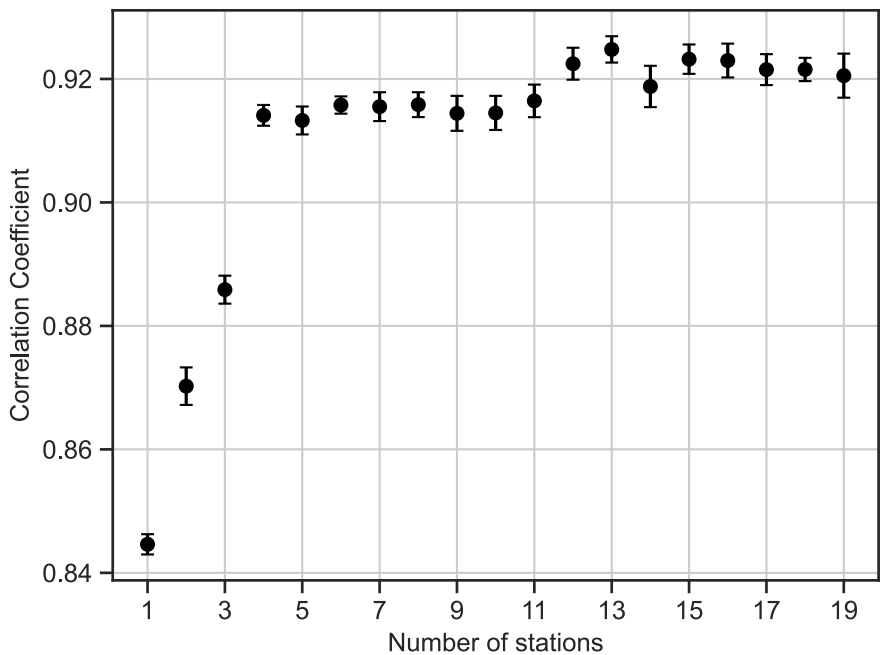

*Figure 6 Pearson correlation coefficient of all the models using different datasets. The error bars indicate the standard deviation of the 20 models' correlation coefficient.*

The outcome of Principal Component Analysis (PCA) associated with the wind data (Figure 7), shows that there was a significant increase in variance explained by the first 4 components, followed by a relatively consistent increase with the inclusion of additional wind stations. The absence of a plateau indicates that each additional component contributes to explaining the dataset's variability. Furthermore, the analysis revealed that more than 35 components are required to account for 99% or more of the variance, which closely aligns with the total number of features employed (38). It is worth noting that the PCA components were tried as the input feature instead of the original stations, but this did not improve the model's results. Accordingly, the original number of features was used as there is no significant benefit from transforming and reducing them using the PCA. These findings affirm that most wind stations, even those far from the simulation location, provide unique information and are not redundant. However, it is important to note that not all stations hold equal importance in predicting wave heights at the specific location of interest.

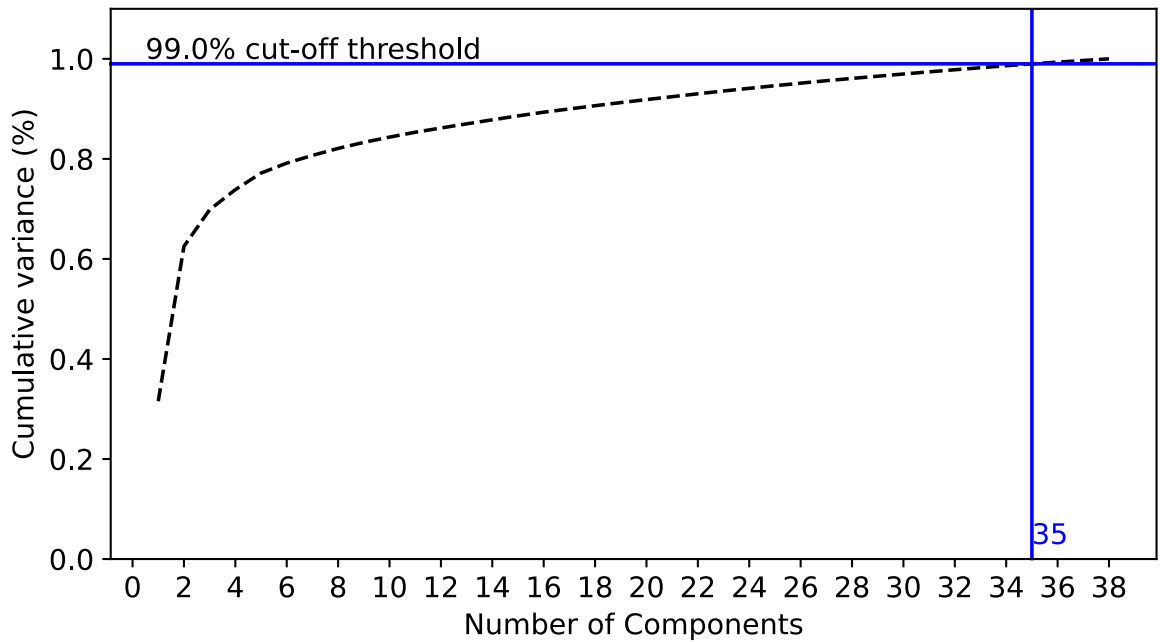

*Figure 7 Principal component analysis (PCA) results for the wind stations. The y-axis shows the percentage of variance explained by the corresponding number of features.*

Grid-search results for optimum kernel sizes and numbers of filters for models with varying numbers of wind stations are shown in Table 3. The kernel size ranged from 1 to 10, with 1 being used for the 1-station model and 10 for the 18-station model. For the other models, the optimum kernel size increased linearly from 1 to 10 with increasing the number of stations. Similarly, the optimum number of filters increased linearly with the number of stations, spanning from 13 to 58. Specifically, the 1-station model employed 13 filters, while the 19-station model utilized 58 filters. In terms of the lookback period, a duration of 10 hours proved to be the most effective across all models. Furthermore, the addition of more ConvLSTM-1D layers or the utilization of alternative activation functions, other than the default tanh function, did not lead to improvements and, in some cases, resulted in a decline in model performance.

During the model training, different logical arrangements of the wind stations within the 1D grid were examined. One potential arrangement was to place the nearest wind stations to the wave's location in the middle of the grid while alternating the subsequent closest stations to the left and right sides. This arrangement was systematically tested across all datasets, and interestingly, it did not yield any significant changes in the model results. These findings highlight the model's insensitivity to station arrangement, as long as the chosen arrangement follows a logical pattern.

*Table 3 Optimized values for the hyperparameters of the different models.*

| Number of wind stations | Kernel size | Number of filters | Lookback period (hours) |
|---|---|---|---|
| 1 | 1 | 13 | 10 |
| 2 | 2 | 16 | 10 |
| 3 | 2 | 18 | 10 |
| 4 | 3 | 21 | 10 |
| 5 | 3 | 23 | 10 |
| 6 | 4 | 26 | 10 |
| 7 | 4 | 28 | 10 |
| 8 | 5 | 31 | 10 |
| 9 | 5 | 33 | 10 |
| 10 | 6 | 36 | 10 |
| 11 | 6 | 38 | 10 |
| 12 | 7 | 41 | 10 |
| 13 | 7 | 43 | 10 |
| 14 | 8 | 46 | 10 |
| 15 | 8 | 48 | 10 |
| 16 | 9 | 51 | 10 |
| 17 | 9 | 53 | 10 |
| 18 | 10 | 56 | 10 |
| 19 | 10 | 58 | 10 |

Figure 8 and Figure 9 present the ensemble learning results for the validation datasets, specifically focusing on the 1-station, 4-station, and 13-station models. The application of a simple ensemble technique in this study yielded modest improvement in the mean square error (MSE) for both models. Notably, the improvement in MSE for the 13-station model reached a plateau after the inclusion of 11 models, whereas the 1-station and 4-station models experienced a slight decline after incorporating 4 and 9 models respectively. Nonetheless, the lowest-error 13-station model maintained a marginally lower MSE of 0.019 $m^2$ compared to 0.022 $m^2$ for the 4-station model and 0.041 $m^2$ for the 1-station model.

In addition to the improvement in the MSE, the ensemble learning approach led to a noticeable reduction in bias for 4-station and the 13-station models (Figure 9). The bias improvement exhibited a plateau for both ensemble models after integrating 8 models, converging to a value close to zero, indicating minimal bias. The bias for the 1-station ensemble model was initially improved for the first 3 models and deteriorated afterwards. This improvement in the MSE and the bias highlight the benefit of the simple ensemble learning technique implemented in this study in harnessing the collective strengths of multiple models, ultimately leading to improving the generalization and the predictive ability of the models.

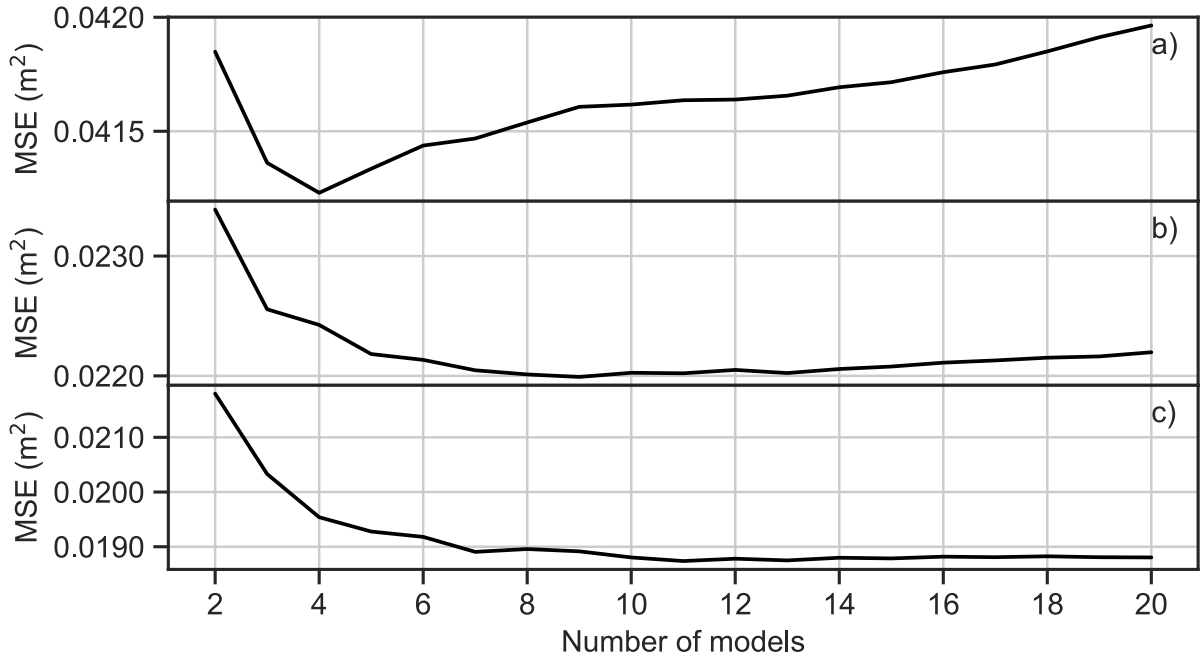

*Figure 8 The MSE for the ensemble learning models. a) shows the results of the 1-station model, b) shows the results of the 4-station model, and c) shows the results of the 13-station model.*

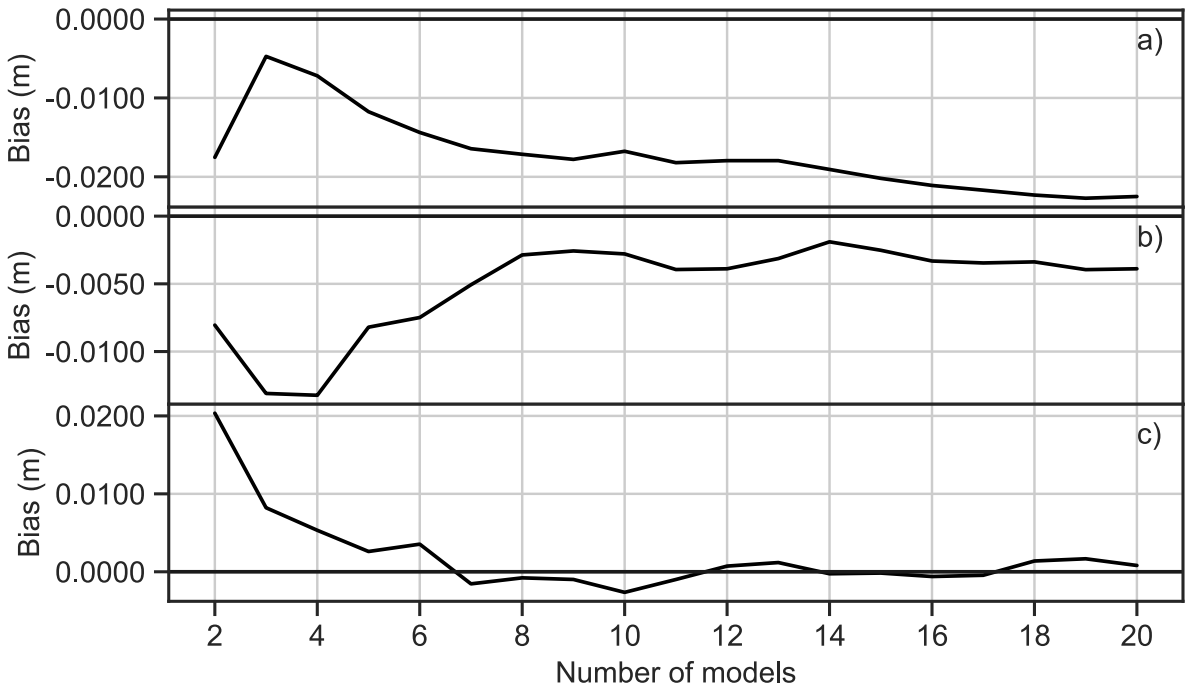

*Figure 9 Bias for the ensemble models. a) shows the results of the 1-station model, b) shows the results of the 4-station model, and c) shows the results of the 13-station model.*

The performance of the best ensemble models for the 1-station, 4-station and 13-station datasets was evaluated on an independent testing dataset consisting of 5 years of wave height data as mentioned earlier. The testing results presented in Figure 10 show a good agreement between the WIS wave heights and the models' predictions. Both models achieve MSE, bias, correlation coefficient (r), and coefficient of determination ($R^2$) values comparable to the values obtained on the validation dataset, which indicates the model's good generalization ability.

Overall, the 13-station model slightly outperformed the 4-station and 1-station model. This can be seen in the superior values of the performance metrics of the 13-station model and in Figure 10, particularly for larger wave heights. It is worth noting that the 4-station and 1-station model offer the advantage of faster training and less missing data as explained earlier. Accordingly, the choice between the three models should depend on the application.

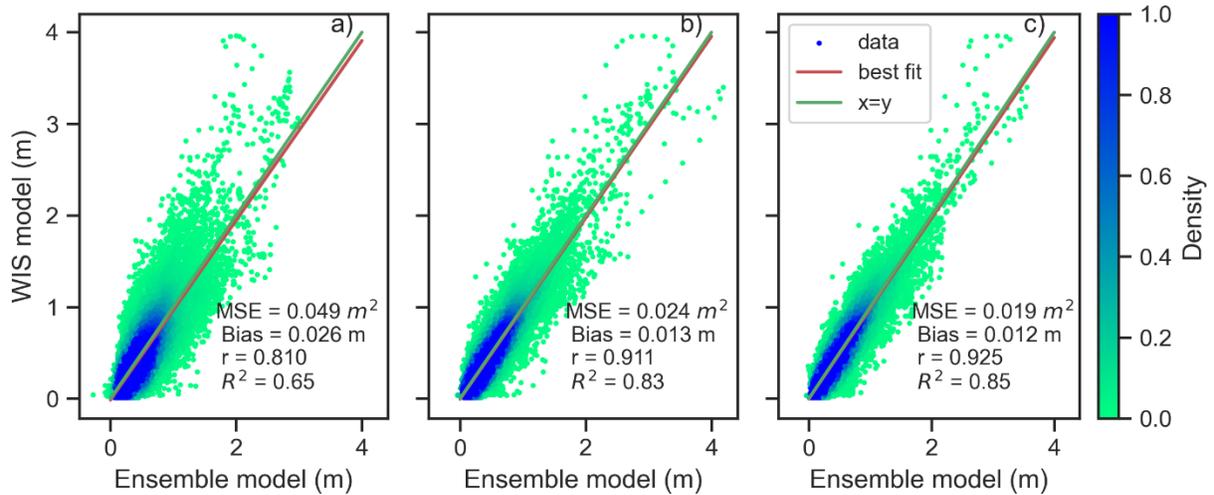

*Figure 10 Ensemble learning results on the testing dataset. a) shows the results of the best 1-station model, b) shows the results of the best 4-station model, and c) shows the results of the best 13-station models. The colormap shows the density of the points.*

Finally, with the 13-station ensemble model determined to be the best overall model of all the models tested and refined, the complete Chicago wave height time series for the period 1949 to 2022 was generated and evaluated (Figure 11). Aside from the fit statistics presented earlier, for the period of overlap with the WIS output used to train and validate the model the resulting time series exhibits very good visual agreement between the model with respect to both the magnitude of the peaks and the timing of the storms. The resulting ML wave simulation successfully extends the hindcast for the Chicago site by 30 years, to create a total hindcast duration of more than 70 years. This wave height time series highlights the ability of the proposed framework in capturing the spatial and temporal correlations between the offshore wind stations and the wave station and to leverage historical wind observations in order to extend wave hindcasts.

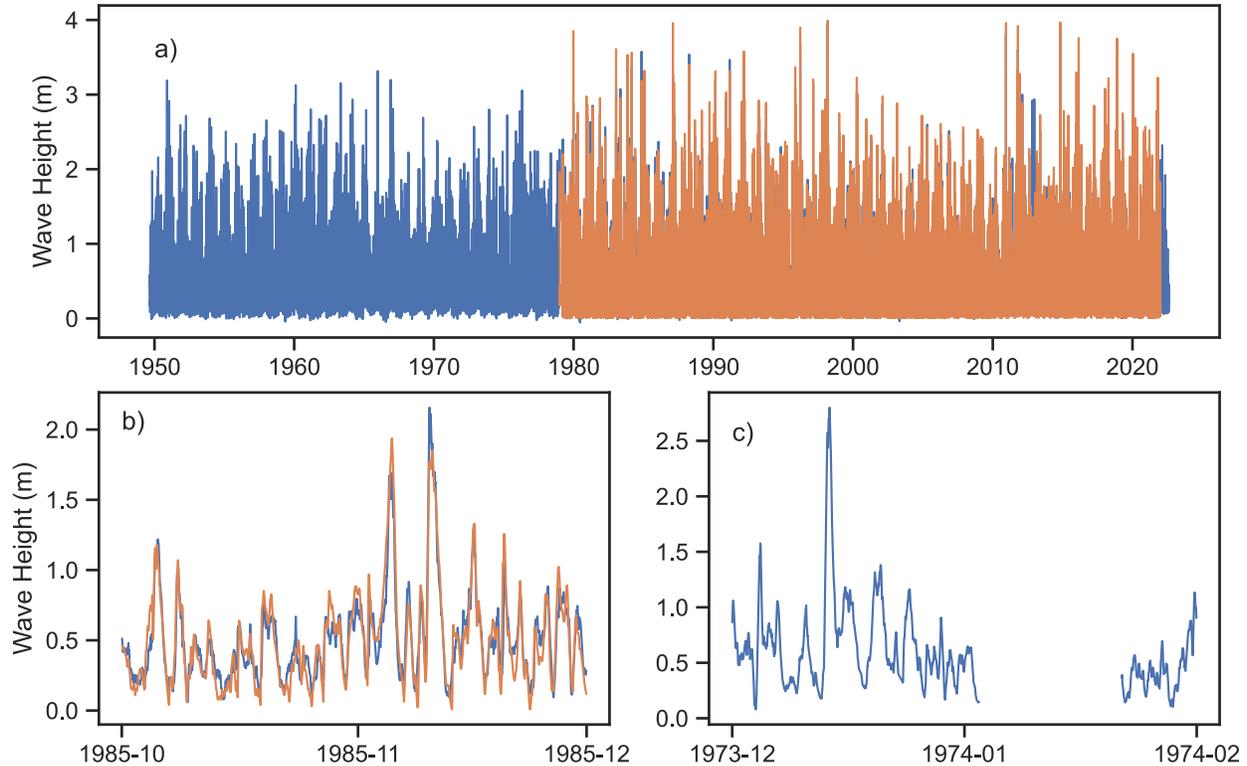

*Figure 11 The generated wave height time series from the 13-station model (in blue), and the WIS calibration/validation data (in orange). a) Shows the full generated wave height time series from 1949 to the present. b) A zoom-in for the time series in the testing dataset. c) Zoom in for part of the hindcasted time series in 1973. The gap in panel number 3 indicated a period with extensive ice cover.*

## Discussion

Similar to previous findings (Ouyang et al., 2023; Song et al., 2022; S. Zhou et al., 2021), this study demonstrates the accurate simulation of wave heights using the ConvLSTM-based machine learning models. This study enhances these efforts by employing the 1D version of ConvLSTM, allowing us to leverage historical non-gridded wind station data. Importantly, our study highlights the capability of ML models to predict wave heights that leverage land-based wind stations located far from the target area of interest for additional accuracy. Furthermore, our results underscore the potential benefits of incorporating multiple wind stations in improving the performance of ML-based wave generation models. The approach developed and tested herein can improve the development of machine learning models for coastal wave prediction in lakes, bays, and semi-enclosed seas.

While the wave height prediction model utilized in this study showed successful results for Lake Michigan, its application to certain ocean coasts may pose challenges. Unlike Lake Michigan, where the majority of waves are wind waves generated by local winds, ocean coasts experience a mix of wind and swell waves. Swell waves are wind waves that have traveled a significant distance from their point of origin, making them difficult to predict solely based on local onshore wind data. Therefore, adapting the model to accurately capture the complex dynamics of ocean coasts, including the presence of swell waves, would require additional considerations and potentially different data sources from the locations of wave generation (Browne et al., 2007).

In the context of duration- and fetch-limited wind waves in Lake Michigan, a short lookback period of less than one day would be expected when modeling waves with nearshore or overlake winds, given standard estimates of wave growth timescales for typical storm intensities and durations seen in the lake (e.g. (Hasselmann et al., 1973)). However, when modeling waves using land stations far from the nearshore location, a longer lookback period may be appropriate, in order to capture the over-land propagation time of the large scale storms generating waves.

While the framework presented in this study relied on ASOS wind data from discrete locations, it can be easily adapted to handle gridded wind data, such as reanalysis datasets. In such scenarios, there is no need for grid construction as the data is inherently gridded. Moreover, employing the 2D version of the ConvLSTM model, as opposed to the 1D version utilized in this study, holds the potential for further improved results. Notably, prior research (Song et al., 2022; S. Zhou et al., 2021) has successfully employed a similar approach, validating the effectiveness of this adaptation. This flexibility enables the incorporation of diverse wind datasets, making it valuable when historical wind data is unavailable.

The inclusion of a time feature as an input for the 1D grid of the model proved beneficial in addressing non-uniform time steps associated with missing wind data. Specifically, models with a larger number of wind stations exhibited noticeable improvements upon incorporating the time feature, whereas models with fewer stations did not demonstrate significant differences. This discrepancy can be attributed to the fact that models with more wind stations tend to have a higher frequency of data gaps compared to those with fewer stations. To tackle non-uniform time steps more comprehensively, advanced techniques such as Neural Ordinary Differential Equations (NeuralODEs) (Chen et al., 2018) or similar algorithms can be employed. However, it should be noted that these methods are computationally intensive and primarily offer advantages when dealing with highly non-uniform wave time series, which is not the focus of this study.

The outcomes of both this study and previous research provide evidence of the advantages offered by ML-based models compared to physics-based models, albeit with some trade-offs. From a computational resource standpoint, ML-based models demonstrate remarkable efficiency in generating wave height time series, with lower computational times than physics-based models (Feng et al., 2020). In the present case, the ML model can enable the generation of wave heights at a specific location without the need to develop an extensive model encompassing the entirety of Lake Michigan. This stands in contrast to physics-based models that often necessitate either modeling the entire lake or utilizing smaller models with offshore wave boundary conditions that are typically unavailable. Lastly, ML models exhibit greater resilience in dealing with input data uncertainty by incorporating ensemble methods or probabilistic predictions, thereby enhancing their applicability in scenarios with data uncertainty (Lakshminarayanan et al., 2017).

Despite the numerous advantages offered by ML models, they are not without limitations. One key drawback is their reliance on extensive training data, unlike physics-based models that can achieve calibration with significantly less data (Stewart, 2019). Additionally, ML models are susceptible to overfitting, wherein they may learn non-physical patterns from the training data, necessitating a larger dataset for validation and testing (Reichstein et al., 2019). Accordingly, in situations where previous wave height data is unavailable, physics-based models remain the sole viable option for generating wave heights solely from wind data, emphasizing their importance in such cases. Additionally, current ML models used for wave generation require training for each new location and cannot generalize to generate wave heights at different locations.

Further research is warranted to enhance the generalizability of ML models across different locations without the need for retraining. Achieving this goal would involve incorporating additional physical features pertaining to the spatial relationship between the wind stations, wave location, and shoreline. By considering factors such as the fetch of the wind from various directions, the model can acquire valuable information that enhances its predictive capabilities. Another promising approach involves non-parametrically representing the wind speed time series by leveraging the distances between the location of interest and the wind station. This technique can offer valuable insights into the spatial variability of wind patterns. Additionally, employing a physics-informed machine learning algorithm with a physics-based loss function holds great potential for reducing the training data requirements while simultaneously improving model generalization and transferability (N. Wang et al., 2022). By integrating domain knowledge into the learning process, such algorithms can effectively leverage both physics-based principles and data-driven techniques, leading to more robust and accurate predictions.

While the model showcased good performance with a remarkably low Mean Squared Error (MSE) on the testing dataset, a closer examination of Figure 11 reveals higher errors associated with extreme wave heights. This discrepancy can be attributed to the scarcity of extreme wave instances within the dataset, which poses challenges for the model to accurately capture and predict such events. Consequently, the significance of comprehensive and lengthy training datasets becomes evident for ML models. Customized loss functions that assigned greater weight to extreme waves were explored during the training process, but this adjustment resulted in increased model bias and compromised overall performance. However, Figure 11 illustrates improvement in modeling extreme waves with the incorporation of additional stations, despite the marginal difference in overall MSE and other performance metrics.(Asch et al., 2022) Overall, however, given that one of the primary motivations in extending the wave hindcast is the desire to better estimate extreme wave properties, more future work is needed on the preferential training of ML models to more accurately simulate extreme waves (Asch et al., 2022).

As mentioned earlier, the presence of ice cover in the Great Lakes and in certain regions of the ocean significantly impacts wave generation, particularly during winter months. This is an important consideration for any wave model when attempting a long-term hindcast. In the case of ML modeling of waves, if ice cover is not considered when training a ML wave model, model performance can be undermined as the model will be trained by periods when winds are high but waves are negligible (due to ice). For the Great Lakes, ice cover data is available from 1973 to present, and this was used to eliminate periods when ice cover was inhibiting waves, so that the model was not trained with these wind/wave/ice conditions. For the period prior to 1973, when the model was used to blindly generate the new period of hindcast, ice cover was taken to be zero (in lieu of any other information), and as such the ML hindcast generated in the present study is an "ice-free" hindcast. This highlights the necessity for further research aimed at also leveraging historical weather data from either land stations or reanalysis products to simulate ice cover, potentially also leveraging ML algorithms. By doing so, the potential of leveraging the extended time series preceding 1973 can be fully harnessed, allowing insights into the long-term wave climatology of the simulation location. Again, it should be restated that the issue of unknown ice cover for historical periods beyond the model training/validation periods is not specific to the machine learning approach, and that physics-based wave models suffer from this same issue.

Nonetheless, the machine learning simulated wave height time series prior to 1973, assuming ice-free conditions, offers valuable insights into the evolving wave climate of Lake Michigan over time. The time series reveals distinct periods characterized by varying storm activity, ranging from periods of small storms to periods of large storms. For instance, the period between 2000 and 2010 exhibits a period of relatively small storm activity, known to coincide with beach rebuilding phases in Lake Michigan, particularly when accompanied by low water levels. Subsequently, after 2010, an increase in storm activity is observed, aligning with a period of rising water levels from 2013 to 2020, which has led to significant erosion. A similar pattern emerges in the time series around 1970, with relatively smaller wave heights compared to adjacent periods. Furthermore, it is notable that the magnitudes of extreme waves prior to 1980 are generally lower than those occurring thereafter. These observations highlight the dynamic nature of the wave climate in Lake Michigan and underscore the influence of various factors, such as storm activity and water levels, on wave behavior and coastal dynamics. As the focus of this current manuscript is the development and refinement of a ML wave prediction model, the full analysis of this wave hindcast time series is left for a separate manuscript.

## Conclusions

In this paper, a new machine learning framework based on the ConvLSTM-1D model was introduced for wave height hindcasting from historical wind records. The framework was applied for hindcasting wave height in Chicago Lake Michigan using wind records from different ASOS wind stations as the input features for the model. Different models with different numbers of stations were trained and tested. While a model forced with only one nearby wind station produced a simulation of reasonable accuracy, it was found that using multiple stations with different distances and orientations from the location of interest of wave prediction significantly improved the results. For the location of interest in this study, the first 4 stations had the greatest effect on improving the model accuracy after which the model performance almost plateaued with the 13-station model being the best. This signifies the importance of choosing the optimum number of stations for such applications.

A simple ensemble learning technique based on the arithmetic summation of different models' results was used in this study. This resulted in an ensemble model with superior performance and better generalization than any individual model for a given number of wind stations. The results also showed the model's ability to capture the spatial and temporal correlation between the wind stations and the wave station, which was shown in its ability to capture both the peak and the time to peak of the storms. The best ensemble model was successfully used to expand the time series of wave height in the Chicago area by adding more than 30 years of wave data starting from 1949 which is almost 70% more than the current available time series for Lake Michigan. Yet, the ice effect was not included before 1973 due to the unavailability of ice cover data, which signifies the importance of expanding the ice cover time series to fully utilize the new wave time series. This extension could also potentially be accomplished using similar machine learning approaches, and an effort to do this is now underway by the authors.

The availability of this extensive time series of wave heights holds significant potential in enhancing our understanding of the wave climate in Lake Michigan. This comprehensive dataset can serve as a valuable resource for studying and analyzing long-term wave patterns, aiding in the development of more robust coastal structures to enhance Lake Michigan's resilience. Moreover, the generated time series enables investigating the correlation between wave characteristics and water level fluctuations in Lake Michigan, contributing to a deeper understanding of the coastal areas within the Great Lakes region.

The proposed framework presents a promising approach for leveraging historical wind records globally to expand wave height time series. By utilizing these extended time series, researchers, engineers, and policymakers can gain valuable insights into the dynamics and trends of wave behavior, empowering evidence-based decision-making for coastal management and infrastructure development. The integration of long-term wave data with other environmental factors and processes will facilitate a holistic understanding of the complex interplay between waves, sea level rise, and coastal dynamics. This comprehensive knowledge will pave the way for more effective strategies in coastal planning, risk assessment, and sustainable development.

## Acknowledgments

This work was supported in part by the Illinois-Indiana Sea Grant College Program, grant number NA18OAR4170082, and the Indiana Department of Natural Resources Lake Michigan Coastal Program, grant number NA20NOS4190036. Hazem Abdelhady also acknowledges support from the Lyles School of Civil Engineering and support from the International Association of Great Lakes Researchers (IAGLR).

## Conflicts of Interest

The authors declare that they have no known competing financial interests or personal relationships that could have appeared to influence the work reported in this paper.

## Declaration of Generative AI Technology in the Writing Process

During the preparation of this work, the authors used ChatGPT in order to improve the writing. After using this tool, the authors reviewed and edited the content as needed and take full responsibility for the content of the publication.